\documentclass[iop]{emulateapj}  

\usepackage{bm}
\usepackage{pslatex}
\usepackage{graphicx}
\usepackage{natbib}

\begin{document}


\title{ Anisotropies of the magnetic field fluctuations at kinetic scales in the solar wind : Cluster observations}


\author{ Catherine Lacombe\altaffilmark{1}, Olga Alexandrova\altaffilmark{1},  and Lorenzo Matteini\altaffilmark{1,2} }


\altaffiltext{1}{LESIA-Observatoire de Paris, PSL Research University, CNRS, UPMC Universit\'e Paris 06, Universit\'e Paris-Diderot, 5 place Jules~Janssen, F-92190 Meudon, France.}

\altaffiltext{2}{Department of Physics, Imperial College London, London SW7~2AZ, UK.}
 


\begin{abstract}



We present the first statistical study of the anisotropy of the magnetic field turbulence in the solar wind between 
1 and 200~Hz,  
{\it i.e.} from proton to sub-electron scales. We consider 93 10-minute intervals of  {\it Cluster}/STAFF measurements.  
We find that the fluctuations $\delta B_{\perp}^2$ are not gyrotropic at a given frequency $f$, a 
property already observed at larger scales ($\parallel$/$\perp$ mean parallel/perpendicular to the average 
magnetic ${\bf B_0}$). This non-gyrotropy gives indications on the  angular distribution of the wave vectors ${\bf k}$:  
at $f <$ 10~Hz, we find that  $k_{\perp}\gg k_{\parallel}$, mainly in the fast wind; at $f >$ 10~Hz,  fluctuations 
with a non-negligible $k_{\parallel}$ are also present. We then consider the anisotropy ratio  
$\delta B_{\parallel}^2/\delta B_{\perp}^2$, which is a measure of the magnetic compressibility of the fluctuations. 
This ratio, always smaller than 1, increases with $f$. It reaches a value showing that the fluctuations are more or less isotropic 
at electron scales, for $f \geq 50$~Hz. From 1 to 15-20~Hz, there is a strong correlation between the observed 
compressibility and the one expected for the kinetic Alfv\'en waves (KAW), which 
only depends on the total plasma $\beta$.  For $f > $  15-20~Hz, the observed 
compressibility is larger than expected for KAW; and it is stronger in the slow wind: this could be an indication 
of the presence of a slow-ion acoustic mode of fluctuations, which is more compressive and 
is favored by the larger values of the electron to proton temperature ratio generally observed in the slow wind.

\end{abstract}

\section{Introduction}
For tens of years, the three orthogonal components of the magnetic field fluctuations have been observed in the solar wind, as 
functions of the frequency $f$ in the frame of a single spacecraft. Their frequency spectra display two kinds of anisotropy. 
A first anisotropy is only related to the direction of the average magnetic field ${\bf B_0}$: there is less energy in the 
compressive fluctuations  $\delta B_{\parallel}^2(f)$ parallel to ${\bf B_0}$ than in the transverse 
fluctuations $\delta B_{\perp}^2(f)$ perpendicular to ${\bf B_0}$. The other anisotropy is related to the direction of ${\bf 
B_0}$ and to the radial direction ${\bf R}$, 
which is close to the direction of the expansion velocity of the solar wind: the fluctuations $\delta B_{\perp}^2(f)$ are 
non-gyrotropic (non-axisymmetric with respect to  ${\bf B_0}$), with less energy in the ${\bf x}$ direction perpendicular to ${\bf 
B_0}$ in the plane $({\bf R},{\bf B_0})$ than in the  ${\bf y}$ direction, perpendicular to both ${\bf B_0}$ and ${\bf R}$ 
\citep[see e.g.,][]{bruno13,oughton15}. 
These two kinds of anisotropy have been observed at large scales, in the inertial 
range  of the magnetic turbulence  
\citep{belcher71, bieber96, wicks12}. 
The fluctuations are still anisotropic and non-gyrotropic at proton scales (around  [0.1, 1]~Hz)  and at sub-proton scales 
up to 10~Hz, a range sometimes called {\it dissipation range}  
\citep{leamon98, turner11}.
 In the present work, we study the anisotropies of $\delta {\bf B}$ from 1~Hz to about 200~Hz, in a sample of 93 
intervals of 10 minutes. 
Indeed, the nature of the turbulent fluctuations in the solar wind is still under investigation, and a study of their anisotropies 
can throw some light on this question: previous works have shown that the anisotropy $\delta B_{\parallel}^2/\delta 
B_{\perp}^2$ gives indications 
on the nature of the dominant mode of the fluctuations, while the non-gyrotropy of $\delta B_{\perp}^2$ 
gives indications on the directions of the wave vectors ${\bf k}$, {\it i.e.} the anisotropy of the ${\bf k}$-distribution. 

The general shape $P({\bf k})$ of the ${\bf k}$-distribution can be deduced from the non-gyrotropy of $\delta B_{\perp}^2(f)$, 
using the fact that fluctuations with different ${\bf k}$ 
contribute to the reduced spectral density observed at a same $f$. If $P({\bf k})$ 
is gyrotropic, a {\it 2D} turbulence  ({\it i.e.} $k_{\perp} \gg k_{\|}$) will be observed with a non-gyrotropic 
$\delta B_{\perp}^2(f)$ in the satellite frame; for a {\it slab} turbulence ({\it i.e. }$k_{\perp} \ll k_{\|}$), 
$\delta B_{\perp}^2(f)$  will be gyrotropic in the satellite frame 
\citep{bieber96}.
Previous observations in the inertial range have shown that the ${\bf k}$-distribution is mainly 2D. 
However, the slab component is also present and its proportion increases when $f$ reaches  
the proton scales [0.1, 1]~Hz 
\citep{bieber96,leamon98, hamilton08}.

The anisotropy of the ${\bf k}$-distribution of the fluctuations has also been estimated by a comparison of the 
correlation lengths parallel and perpendicular to ${\bf B_0}$, with a single spacecraft, assuming a stationary and 
gyrotropic ${\bf k}$-distribution in the MHD range,  
below $10^{-3}$~Hz \citep{matthaeus90} 
and below $10^{-2}$~Hz \citep{dasso05}.
Multispacecraft measurements also allow a comparison of the correlation lengths, parallel and perpendicular to 
${\bf B_0}$, in the inertial range, still assuming a gyrotropic ${\bf k}$-distribution. \citet{osman07}, 
with {\it Cluster} observations, 
find an indication that the ${\bf k}$-distributions in the inertial range are more 2D for the compressive fluctuations 
$\delta B_{\parallel}^2$ than for the transverse fluctuations $\delta B_{\perp}^2$. 

Using data from the four {\it Cluster} spacecraft, \citet{chen10} 
find that $\delta B_{\parallel}^2$ and $\delta B_{\perp}^2$ 
have a ${\bf k}$-distribution with $k_{\perp} > k_{\parallel}$, below 10~Hz.  Using the $k$-filtering technique, 
\citet{sahraoui10} find  $k_{\perp} \gg k_{\parallel}$ up to  2~Hz.  
Still using the $k$-filtering technique below 1~Hz, for 52 intervals of {\it Cluster} data, \citet{roberts15b} 
find $k_{\perp} \gg k_{\parallel}$; except in four intervals with a relatively fast wind,  
where quasi-parallel wave vectors are also present above 0.05~Hz, up to 0.3~Hz where they disappear \citep{roberts15a}. 
These quasi-parallel wave vectors correspond to Alfv\'en Ion Cyclotron waves (AIC) which are 
usually observed in this frequency range, in a fast wind \citep{jian14}.
At proton scales, in the range [0.1,2.5]~Hz, \citet{perrone16} 
find that the magnetic fluctuations in a slow wind are 
mainly compressive coherent structures,  with  $k_{\perp} \gg k_{\parallel}$. At the same scales, but in a fast wind, 
\citet{lion16} find Alfv\'enic coherent structures with  $k_{\perp} \gg k_{\parallel}$, and quasi-monochromatic AIC waves 
with $k_{\|} \gg k_{\perp}$, superimposed on a non-coherent and non-polarised component of the turbulence. 
In these analyses, as well as in \citet{perschke13}, 
there is no clear mention of a non-gyrotropy of the wave vector 
energy distribution in the solar wind. 

The second kind of anisotropy,  the ratio $\delta B_{\parallel}^2 / \delta B_{\perp}^2$, is a measure of the compressibility of the 
magnetic fluctuations. In the inertial range, as well as at proton scales, the compressibility increases when the proton 
$\beta_p$ factor increases \citep{smith06, hamilton08}. 
$\beta_p$ ($\beta_e$) is the ratio between the proton (electron) thermal energy and the magnetic energy. 
The compressibility also increases when $f$ increases, from the inertial range to proton scales 
\citep{hamilton08, salem12} 
and from 0.3 to 4~Hz \citep{podesta12}. 
At sub-proton scales, the compressibility still increases with $\beta_p$  up to 10~Hz \citep{alexandrova08a} 
and still increases with $f$ up to 100~Hz \citep{kiyani13}. 
We call  $d_p$  and $d_e$ the proton and electron inertial lengths, $\rho_p$ and $\rho_e$ the proton and electron gyroradii, 
calculated with the temperatures perpendicular $T_{p\perp}$ and $T_{e\perp}$ perpendicular to ${\bf B_0}$. 

Even if isolated, coherent and non-linear structures are present in the solar wind \citep{perri12, perrone16, lion16}, 
the comparison of observed dimensionless ratios of the fluctuating fields and plasma quantities (transport ratios) 
with the linear properties of the plasma wave modes can give indications, 
in a first approximation, on the nature of the dominant type of the fluctuations \citep{gary92, krauss94, denton98}. 
This usual modelling of the turbulence as a superposition of linear waves 
is discussed by \citet{klein12} and \citet{tenbarge12} 
who consider Alfv\'en, fast, whistler and slow modes from the MHD to the kinetic range. 

The slow mode is frequently neglected because it is strongly damped, except when the 
electron to proton temperature ratio $T_e/T_p$ is larger than 1. However  {\it Wind} observations at large scales 
($k \rho_p < 0.05$), in the inertial range, show that  the 
anticorrelation between the density and the compressive magnetic fluctuations is typical of 
quasi-perpendicular slow modes;  kinetic slow waves may then be cascaded as passive fluctuations by Alfv\'enic 
fluctuations, and thus exist at proton scales \citep{howes12, klein12}. 
\citet{narita15} underline how difficult it is to distinguish between kinetic slow modes and  kinetic Alfv\'en waves (KAW), 
mainly in a high-beta plasma. 

\citet{salem12} compare the observed compressibility with the compressibility expected for whistler waves or for KAW, and 
find that the compressibility up to 1~Hz can be explained by KAW-like  fluctuations, with nearly perpendicular wave vectors. 
Quasi-perpendicular KAW are generally found up to proton scales, {\it i.e.} up to 1 to 3~Hz 
\citep{sahraoui10, he12, podesta12, podesta13, roberts15b}. Conversely, 
\citet{smith12} find that KAW cannot be the only component below 1~Hz, at least when $\beta_p$ is larger than 1. 
Comparing the spectra of the density and of the magnetic field fluctuations measured 
on the ARTEMIS-P2 spacecraft, 
\citet{chen13} find KAW between 2.5 and 7.5~Hz, {\it i.e.}  $k \rho_p \simeq$ 5 to 14.

Calculations of \citet{podesta10} 
show that the linear collisionless electron Landau damping prevents KAW  
to cascade to the electron scales, so that the energy cascade to these scales must be supported by wave modes other than 
the KAW mode. On the other hand, analytical calculations and numerical simulations of the turbulence at sub-proton  
scales show that a cascade characterized by KAW-like properties can be sustained between proton and electron scales
\citep{howes11, tenbarge13, franci15, schreiner17} and can explain the shape of the spectra observed by  
\citet{alexandrova12}, controlled by the electron gyroradius.  

With 2D kinetic simulations, \citet{camporeale11b} 
analyse the dispersion relation and the electron compressibility ({\it i.e.} the ratio between the electron density and the 
magnetic field modulus of the fluctuations) up to $k \rho_e =$ 1 or more, for $\beta_p = \beta_e = 0.5$. They find 
that the electron compressibility of the simulated fluctuations 
is too small to be related to slow-ion acoustic waves, and much too large to be related to whistler waves, while KAW are 
damped. They conclude that the fluctuations are probably a mixture of different modes. 

 In this paper, we consider the two types of anisotropy discussed above: (i) the non-gyrotropy of 
$\delta {\bf B}_{\perp}$ and (ii) the compressibility of the magnetic fluctuations $\delta B_{\parallel}^2 / \delta B_{\perp}^2$. 
After a presentation of the {\it Cluster/STAFF} instrument (section 2) and of the data selection procedure (section 3), 
we study the non-gyrotropy of $\delta {\bf B}_{\perp}$. In section 4,  we compare our observations, from 1~Hz to  50~Hz or more,  
in the fast wind and in the slow wind,  with the calculations of \citet{saur99}: 
this gives us indications on the 
${\bf k}$-distribution, assumed to be gyrotropic  in the plasma frame. Note that the values of the field-to-flow angle $\theta$  
are near $90^{\circ}$ in our sample, otherwise {\it Cluster} would be magnetically connected to the Earth's bow shock. 
In section 5, we study the compressibility of the magnetic fluctuations. 
We show how the compressibility 
depends on the plasma parameters in our sample. Then, we compare our observations with the predictions for KAW. We show 
that  the observed magnetic compressibility  agrees with the KAW compressibility up to 15-20~Hz, which corresponds to 
$k d_e \simeq 0.3$ in our sample (where the electron inertial length $d_e$ is about 1 to 3 km). 
Above 20~Hz, the magnetic fluctuations are more compressive than KAW, and more compressive in the slow wind than 
in the fast wind. It is well-known that the electron to proton temperature ratio $T_e/T_p$ is larger in the slow wind. 
A larger $T_e/T_p$ increases the damping of KAW for several values of 
$\beta_p$ and $k$ \citep{schreiner17} 
but reduces the damping of the slow-ion acoustic mode;  so that we cannot exclude 
the presence of this last mode for $f > $ 20~Hz, {\it i.e.} for $k d_e \ge$ 0.3, a mode which would be less damped and more 
compressive than the KAW mode.

\section{Instruments and data}
The present study relies on data sets from different experiments onboard a spacecraft ({\it Cluster}~1) of the {\it Cluster} fleet.
The Spatio-Temporal Analysis of Field Fluctuations (STAFF) experiment on {\it Cluster} \citep{cornilleau97, cornilleau03} 
measures the three orthogonal components of the magnetic field fluctuations in the frequency range 
0.1~Hz - 4~kHz, and comprises two on board analysers, a wave form unit (SC) and a Spectrum Analyser 
(SA). STAFF-SC provides digitized wave forms, which are projected in the Magnetic Field Aligned (MFA) frame given 
by the FGM experiment every 4 seconds; Morlet wavelet spectra are then calculated 
between 1 and 9~Hz. The Spectrum Analyser builds a 3x3 spectral matrix every 4 s, at 27 frequencies between 8~Hz and 4~kHz. 
This spectral matrix is also projected in the MFA frame: the Propagation Analysis of STAFF-SA Data with 
Coherency Tests (PRASSADCO program) gives the fluctuation properties, direction of propagation, phase and 
polarisation with respect to ${\bf B_0}$ \citep{santolik03}. 
Both experiments, SC and SA, allow to see whether the observed fluctuations are polarised or not. 
In some intervals, circularly-polarised fluctuations are observed, with a direction of propagation near ${\bf B_0}$, at 
frequencies displaying a small or large spectral bump. These  fluctuations are quasi-parallel whistler waves \citep{lacombe14}. 
In the present work we only consider intervals without such quasi-parallel whistler waves.

The WHISPER experiment \citep{decreau97} 
is used to check that {\it Cluster} is in the free solar wind, 
{\it i.e.} that the magnetic field line through {\it Cluster} does not intersects the Earth's bow shock:
there is no electrostatic or Langmuir wave, typical of the foreshock, in our sample. 
Some of the intervals studied by 
\citet{perri09}, \citet{narita10, narita11a, narita11b}, \citet{narita14}, \citet{comisel14}, and \citet{perschke14} 
contain incursions of {\it Cluster} in the foreshock.

The Cluster Science Data System gives the magnetic field ${\bf B_0}$ every 4~s (FGM experiment, \citet{balogh97}), 
the proton density $N_p$, the wind velocity $V_{sw}$  and the proton temperatures $T_{p\parallel}$ and $T_{p\perp}$, 
parallel and perpendicular to ${\bf B_0}$ of the whole proton distribution (CIS 
experiment, \citet{reme97}). 
The electron parameters  given by the Low Energy Electron Analyser of the PEACE experiment \citep{johnstone97} 
are taken from the Cluster Science Archive: we  use the electron  
temperatures $T_{e\parallel}$ and $T_{e\perp}$ of the whole electron distribution. 

\section{Data set}

 
\begin{figure}[!h]
\begin{center}
\includegraphics [width=9cm]{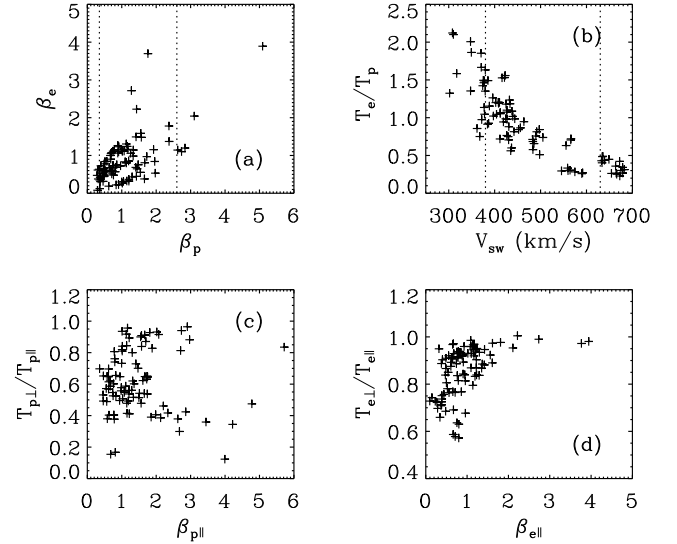}
\caption{Ranges of the solar wind parameters  of our sample. a)  the factors $\beta_p$ and $\beta_e$. 
The dotted lines indicate the ranges of small and large $\beta_p$ considered in Figures 6 and 7d. 
b) The wind speed and $T_e/T_p$. The dotted lines indicate the ranges of slow and fast wind considered in 
sections 4.3 and 5.2. 
c)  $T_{p\perp} / T_{p\parallel}$ is always weaker than 1.
d) $T_{e\perp} / T_{e\parallel}$ is generally weaker than 1. }
\label{fig1}
\end{center}
\end{figure}

We first consider the same sample of solar wind intervals as \citet{alexandrova12},
observed on {\it Cluster}~1 from 2001 to 2005. This sample contains 100  intervals of 10 minutes, when magnetic  fluctuations 
(which are not circularly polarised) are observed above 1~Hz, up to 20 to 400~Hz according to the intensity of the fluctuations.  
These frequencies are above the proton cyclotron 
frequency  $f_{cp} = $  [0.08 - 0.3~Hz] in our sample.  With the Taylor hypothesis 
$2\pi f = \omega \simeq {\bf k} \cdot {\bf V_{sw}}$, where $k$ is the wavenumber of the fluctuations, the 
scales corresponding to the observed frequencies decrease from $k d_e \simeq 0.01$ to $k d_e \simeq 3$
($k d_p \simeq$ 0.4 to 130).

For each 10-minute interval, average spectra are calculated over 150 points (4~s spectra) in the three 
directions of the MFA frame. The 4~s FGM magnetic field  ${\bf B_0}$ can be considered as a {\it quasi-local} mean field, 
giving an average frame valid for fluctuations above $\simeq$ 1~Hz.  It is not really a {\it local} mean ${\bf B_0}$, 
which would be computed with waveforms at each scale: such a computation is impossible with the STAFF-SA data, 
above 1~Hz. A consequence of this use of a {\it quasi-local} mean ${\bf B_0}$, in place of a local field, will be mentioned in section 4.1.

\citet{alexandrova12} studied the spectral shape of the total variance of the fluctuations, summed in 
the three directions (without subtraction  of the background noise). 
We now want to study the anisotropies of the variance of the magnetic fluctuations. It is well known that the variance is 
anisotropic, in the inertial range and at kinetic scales, and that the anisotropies are weaker at small scales 
(see  {\it e.g.}  \citet{turner11},  and references therein; \citet{kiyani13}).
At high frequencies, the Power Spectral Density (PSD) is weaker, and tends towards the instrument 
background noise, which is an additive noise. It is thus necessary to subtract the background noise, not 
only for the total 3D magnetic 
variance but also for the 1D variance in each direction. This subtraction has been done as explained in 
Appendix.  As a result, 7 intervals among 100 have been withdrawn from the sample because  their signal-to-noise ratio
was smaller than 3 at 20~Hz and above. We thus obtain 93 intervals with 1D and 3D spectra intense enough up to 20~Hz or more, 
after subtraction of 1D and 3D background noise. The importance of the noise subtraction is illustrated in 
Figure 4 of \citet{howes08}.

During these intervals in the free solar wind, the {\it Cluster} orbit implies that 
the average magnetic field ${\bf B_0}$ makes a large angle with 
the solar wind velocity: the sampling direction is far from the ${\bf B_0}$ direction. 
The acute angle $\theta$ between ${\bf  B_0}$ and ${\bf  R}$, which is close to the field-to-flow angle, is always larger 
than $52^{\circ}$ in our sample; its average value is $78^{\circ}$. 

 \begin{figure}[!h]
\begin{center}
\includegraphics [width=9cm]{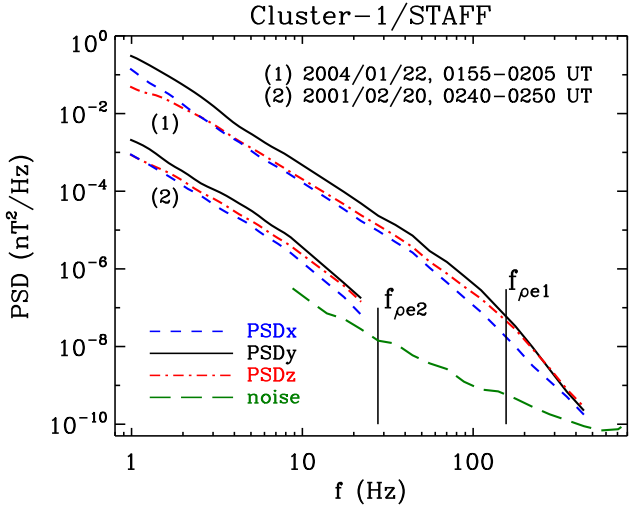}
\caption{The most intense, and one of the 
weakest spectra of our sample.  
The green long-dashed line gives the average background noise $bn_{1D}(f)$ on a search coil (see Appendix). The two vertical solid 
lines give the frequency $f_{\rho_e} = V_{sw} / (2 \pi \rho_e)$ corresponding to the electron gyroradius scale for the two spectra, (1) and (2)}.
\label{fig2}
\end{center}
\end{figure}

Figure 1 gives the ranges of the plasma properties which 
can be found in our data set. It displays scatter plots of the beta factors $\beta_e$ and $\beta_p$, of 
the solar wind speed $V_{sw}$, of the electron to proton temperature ratio $T_e/T_p$, and of the anisotropy ratios of the proton 
and electron temperature, $T_{p\perp} / T_{p\parallel}$ and $T_{e\perp} / T_{e\parallel}$. 
$\beta_p$ varies between 0.28 and 5.1, and $\beta_e$ between 0.08 and 3.9 (Fig. 1a). 
$T_e/T_p$ varies between 0.22 and 2.2, and $V_{sw}$ between 300 and 690 km/s (Fig. 1b).  $T_{p\perp} / T_{p\parallel}$ 
varies between 0.12 and 0.97 (Fig. 1c), and $T_{e\perp} / T_{e\parallel}$ between 0.57 and 1 (Fig. 1d). 
Within these ranges, the plasma is expected to be sufficiently far from the thresholds of proton and electron 
kinetic instabilities 
\citep[e.g.,][]{Hellinger_al_2006, Stverak_al_2008, Matteini_al_2013, Chen_al_2016}, 
at least in the majority of the intervals analysed. We then exclude that a possible wave activity associated 
to these processes may significantly affect the results of this work.

\section{Observations, and wave vector distribution}

\subsection{Spectra and anisotropies}

\begin{figure}[!h]
\begin{center}
\includegraphics [width=9.cm]{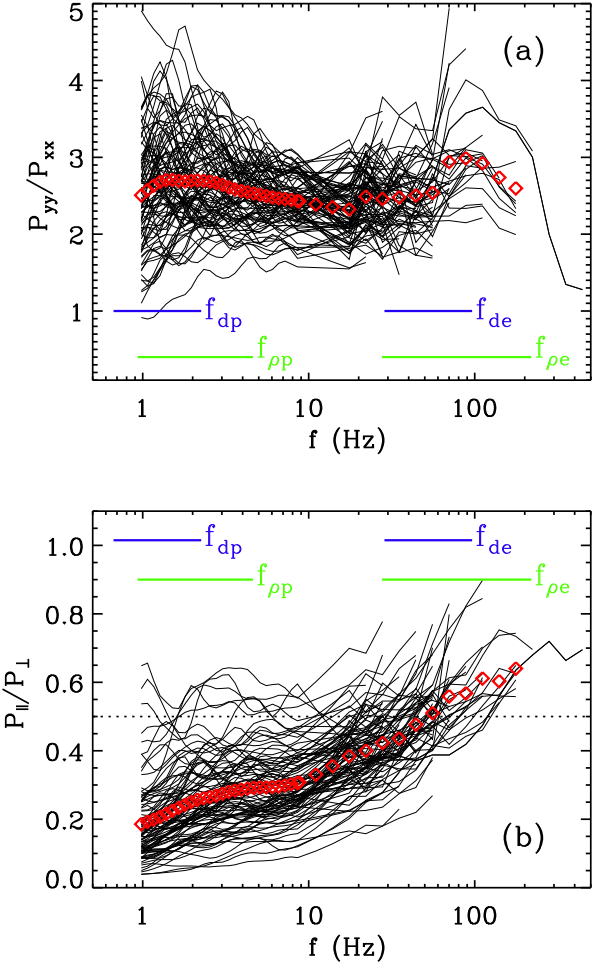}
\caption{Variance anisotropies: ratios between the 10-minute averaged PSD in different directions. The solid lines are for 
the 93 events. The red diamonds are averages of these ratios, at each frequency. 
a) ratio $(P_{yy}/P_{xx})(f)$ between the average PSD in the 
two directions {\bf x} and {\bf y} perpendicular to ${\bf B_0}$. 
b) the magnetic compressibility $(P_{\parallel} / P_{\perp})(f)$, ratio between the compressive PSD and the total 
transverse PSD. In each panel,  the thick horizontal bars give the frequency range of 
$f_{\rho_p} = V_{sw} / (2 \pi \rho_p)$, 
$f_{d_p} = V_{sw} / (2 \pi d_p)$, $f_{\rho_e} = V_{sw} / (2 \pi \rho_e)$ and 
$f_{d_e} = V_{sw} / (2 \pi d_e)$  in the sample.} 
\label{fig3}
\end{center}
\end{figure}

We consider the anisotropic Power Spectral Density (PSD) $P(f)$ of the magnetic fluctuations as a function of 
$f$. The PSD of the transverse fluctuations, perpendicular to ${\bf B_0}$, is $P_{\perp}(f) = P_{xx}(f) + P_{yy}(f)$.  
$P_{yy}$ is the PSD perpendicular to  ${\bf B_0}$ and to ${\bf R}$, 
and $P_{xx}$ is the PSD perpendicular to ${\bf B_0}$ in the plane $({\bf R},{\bf B_0})$. 
We call compressive the fluctuations parallel to ${\bf B_0}$ {\it i.e.} to ${\bf z}$, $P_{\parallel} = P_{zz}$. We call 
$P_{3D}(f)$ the sum of the fluctuations in the three directions $P_{3D} = P_{\parallel} + P_{\perp}$. 
Two spectra of our sample are shown in Figure 2. For each spectrum, the solid line gives  $P_{yy}$, the dashed line 
$P_{xx}$ and the dash-dot line $P_{zz}$. We see that $P_{yy}$ is the most intense spectrum and that $P_{xx}$ is 
generally the least intense. $P_{zz}$ is close to $P_{xx}$ at the lowest frequencies, and 
close to $P_{yy}$ at the highest frequencies.

Figure 3a displays the anisotropy of the variance of the fluctuations in the two directions ${\bf x}$ and ${\bf y}$ 
perpendicular to ${\bf B_0}$. The 93 solid lines are 
the ratios of the 10-minute averages of the PSD in these two directions, anisotropy ratios as functions 
of $f$ (see a discussion about different averaging methods in \citet{tenbarge12}).
The red diamonds are average values of these anisotropy ratios over the 93 events, at each frequency. 
$P_{yy}/ P_{xx}$ is always larger 
than  1. This implies that the transverse fluctuations are not gyrotropic at a given frequency in the 
spacecraft frame. The fact that the non-gyrotropy of  $P(f)$ could be 
compatible with a gyrotropic  wave vector energy distribution has been suggested by \citet{alexandrova08b}. 
By analytical calculations and numerical simulations, 
\citet{turner11} have shown that this suggestion was 
valid for the magnetic fluctuations in the solar wind, from the inertial range to 10~Hz, if the frequency $f$ is only 
due to the sampling along $V_{sw}$ (Taylor hypothesis) and 
if the modulus $q$ of the spectral index of the fluctuations is larger than 1.

The compressibility $P_{\parallel} / P_{\perp}$ of the 93 events, observed as a function of $f$, is given in Figure 3b: 
the average value of 
$P_{\parallel} / P_{\perp}$ at each frequency is shown by red diamonds. The compressibility increases when $f$ increases. 
The Figure 10 of \citet{tenbarge12} 
illustrates how the compressibility depends on the chosen local ${\bf B_0}$ 
field: the compressiblity calculated with a field averaged over 1~hour can be 2 to 3 times larger than the compressibility 
calculated with a local field. We thus think that our {\it quasi-local} field (4~s average) can lead to a slight overestimation 
of the compressibility at the highest frequencies.

According to Figure 3 (see the horizontal bars in each panel), the frequencies $f_{\rho_p}$ corresponding to 
the proton gyroscale ($k \rho_p = 1$) 
are generally found between 1 and 4~Hz, and the frequencies  $f_{d_p}$ between 0.7 and 2~Hz. For the electron scales, 
$f_{\rho_e}$  is mainly found between 30 and 200~Hz, and  $f_{d_e}$ between 30 and 100~Hz.

In the following subsection, we shall look for wave vector distributions consistent with the non-gyrotropy in the 
frequency domain displayed in Figure 3a. The other anisotropy, the compressibility $P_{\parallel} / P_{\perp}$ (Fig. 3b), 
will be studied in section 5. 

\subsection{Angular distributions of the  wave vectors}

\citet{saur99} consider four models for the wave vector angular distribution in the solar wind, and the resultant 
PSD anisotropies in the frequency domain, assuming that the fluctuations are stationary and frozen-in (Taylor hypothesis). 
The four distributions are: 

a) a slab symmetry with respect to ${\bf B_0}$ {\it i.e.} ${\bf k}$  strictly parallel to ${\bf B_0}$.

b) a gyrotropic 2D distribution, with ${\bf k}$  strictly perpendicular to ${\bf B_0}$.
The corresponding fluctuations are in planes which contain ${\bf B_0}$, and can make any angle with it, with a 
component $P_{\parallel}$ parallel to ${\bf B_0}$ and a component $P_{\perp}$ perpendicular to ${\bf B_0}$. 

c) an isotropic  wave vector distribution. 

d) a slab symmetry with respect to the radial direction {\it i.e.} ${\bf k}$ strictly parallel to ${\bf R}$.

\noindent Equations (33) to (43) of  \citet{saur99} 
give the values of  $P_{xx}(f)$, $P_{yy}(f)$ and $P_{zz}(f)$ for the 
four above distributions, as functions of $f$,  $\theta$ and $(P_{\parallel}/P_{\perp})(f)$. The PSD are power laws $\simeq f^{-q}$. 
The spectral index $q$ of the fluctuations is assumed to be the same in the three directions $x$, $y$ and $z$, 
and constant over the whole frequency range. 

The equations of 
\citet{saur99} are valid at  MHD scales, larger than  
$\rho_p$ and $d_p$. However, at smaller scales, the Hall term starts to show its effects in the induction equation. 
As clearly shown by \citet{kiyani13} \citep[see also the detailed derivation of][]{schekochihin09}, 
the Hall term implies interactions between the compressive $P_{zz}$ and 
transverse $P_{\perp}$ fluctuations, and an enhancement of the magnetic compressibility. 
Thus, the MHD terms $P_{zz}/P_{yy}$ and  $P_{zz}/ P_{xx}$ given by \citet{saur99} 
are no more valid at proton and electron scales. 
The term $P_{yy}/ P_{xx}$, which only involves transverse fluctuations, can be considered as less affected by the Hall effect: 
it does not depend on $P_{\parallel}/P_{\perp}$. 
It is thus able to give indications on the wave vector distribution at scales smaller than the MHD scales. 


 \begin{figure}[!h]
\begin{center}
\includegraphics [width=9cm]{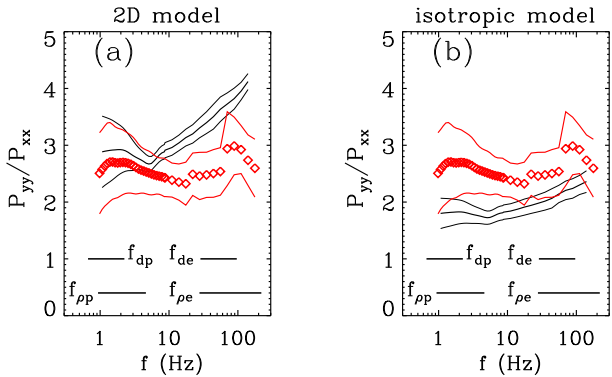}
\caption{For $(P_{yy}/P_{xx})(f)$, comparison of the observations with two models (Eqs. 2 and 3). 
Red diamonds: the observed average anisotropies, given by diamonds in Figure 3a; the error interval at each frequency, 
related to the standard deviation over the 93 intervals, is given by the red lines. 
Thick black lines: the average anisotropy calculated for two models of the angular distribution of the wave vectors;
a) the 2D model (Eq. 2);  b)  the isotropic model (Eq. 3). The thin black lines give the standard deviation for each model, 
related to the dispersion of the spectral index $q$ over the 93 intervals.}
\label{fig4}
\end{center}
\end{figure}

The anisotropy $P_{yy}/ P_{xx}$ can be calculated with the formulas of 
\citet{saur99} for the four above distributions. For the slab symmetry with respect to $B_0$, 

\begin{equation}
(P_{yy} / P_{xx})(f) = 1 .
\end{equation}

\noindent This distribution gives a gyrotropic variance, which is not observed. For the 2D distribution, 
$P_{yy}(f)$ and  $P_{xx}(f)$ depend on $(P_{\parallel}/P_{\perp})(f)$ in the same way, so that 
their ratio does not depend on it:

\begin{equation}
(P_{yy} / P_{xx})(f) = q ,
\end{equation}

\noindent where $q$ is the spectral index of $P_{xx}$ and $P_{yy}$. 
For the isotropic distribution, $P_{yy}(f)$ and  $P_{xx}(f)$ do not depend on $(P_{\parallel}/P_{\perp})(f)$:

\begin{equation}
(P_{yy}/ P_{xx})(f) = \frac {(1+q)/2}{{\rm cos}^2\theta (1+q)/2 +{\rm sin}^2\theta} 
\end{equation}

\noindent  where $\theta$ is the field-to-flow angle.  For the slab symmetry with respect to ${\bf R}$,

\begin{equation}
(P_{yy}/ P_{xx})(f) = 1/{\rm cos}^2\theta, 
\end{equation}

\noindent In Figure 4, the observations of  $P_{yy}/ P_{xx}$ are compared with two models (Eqs. 2 and 3). 
The calculated anisotropies depend on the observed spectral index $q$. To get an average 
value $<q>$ of the spectral index at each frequency, we use successive averaging processes: we first calculate the local slope 
(linear fit over three contiguous frequencies) of the total spectrum $P_{3D}$ of each event of our sample; then, 
we smooth this local slope over 11 contiguous frequencies for each event; finally, we 
average the slope over the 93 events, and find $<q>$.  

In Figures 4a and 4b, the observed average anisotropy (red diamonds)  $P_{yy}/ P_{xx}$ shown in Figure 3a is drawn as a function 
of $f$. As the dispersion of $P_{yy}/ P_{xx}$ for the 93 events is large (Fig. 3a), the two red lines of Figures 4a and 4b 
give the error interval (average $\pm$ the standard deviation) at each frequency. 
Following Eq. (2), $P_{yy}/P_{xx}$ 
for the 2D distribution is equal to the spectral index. Thus, in Figure 4a, the thick black solid line for the 2D ${\bf k}$-distribution 
shows  the observed average $<q>$: it varies between 2.7 and 2.9 below 6~Hz; then it increases up to 4 when $f$ increases. 
This regular increase of $<q>$ is due to the exponential shape of the spectra \citep{alexandrova12}. 
The thin black lines give the standard deviation for $<q>$ over the 93 events, {\it i.e.} the dispersion of the 
2D model itself.
Taking into account the standard deviation of the observed ratio $P_{yy}/P_{xx}$ (red lines), it is 
clear that the observations from 1~Hz to 6~Hz are close to the ratio for a 2D ${\bf k}$-distribution. 
Above 10~Hz  (at sub-proton and electron scales) the observations are further from the 2D model. 

In Figure 4b, the observations are compared with the isotropic model (with its standard deviation). 
Below 10 ~Hz, the observations are far 
from the isotropic model. Above 10~Hz, the observations are near the ratio for an isotropic ${\bf k}$-distribution. At any 
frequency, the slab model  (Eq. 1) is far from the observations. The ratio (Eq. 4) for the model with a slab symmetry 
with respect to ${\bf R}$  is not drawn in Figure 4 because it is generally larger than 10. 
We conclude that the 2D model is better below 6~Hz, 
and the isotropic model better above 10~Hz.
However, this conclusion relies on the hypothesis of \citet{saur99} 
that the spectral index $q$ is constant over the whole frequency range: this is not 
exact above about 10~Hz (Fig. 4a). Our results are just an indication, given by the non-gyrotropy of the PSD, 
that the $k$-distribution tends to be 2D below 6~Hz, and more isotropic above 10~Hz. 

Several authors assume that the turbulence is in critical balance, {\it i.e.} the nonlinear cascade rate is 
of the order of the linear frequency. Such a model implies relations between 
$k_{\parallel}$ and $k_{\perp}$, {\it i.e.} privileged directions of the wave vectors, with more and more energy 
in $k_{\perp}$ when the wavenumber increases 
\citep[see Fig. 1 of][]{howes08}. It can be valid from the inertial 
range to the kinetic range \citep{goldreich95, howes08, schekochihin09, tenbarge13}. 
In our sample, the critical balance model could be valid at low frequency (below 6~Hz) where the wave vectors are 
quasi-perpendicular; but it is probably less valid at higher frequency, where the ${\bf k}$-distribution is more isotropic, 
{\it i.e.} with relatively more energy in oblique or parallel wave vectors  when the wavenumber increases. 


\begin{figure}[!t]
\begin{center}
\includegraphics [width=9cm]{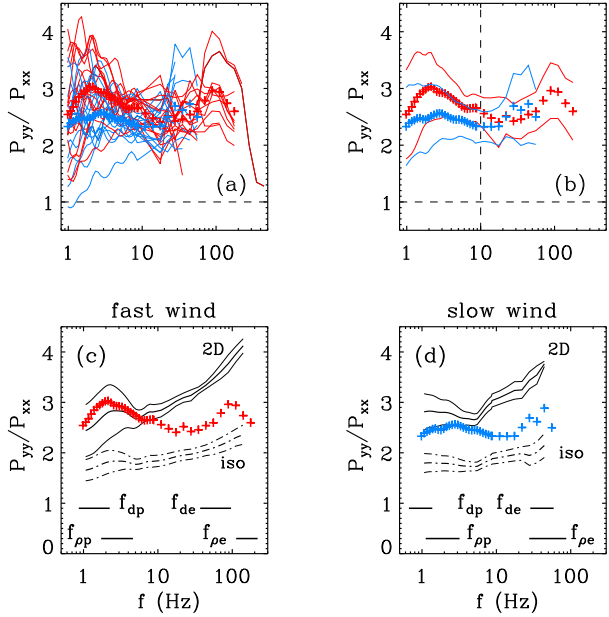}
\caption{Ratios $(P_{yy}/P_{xx})(f)$ for two samples: a sample (16 events) with $V_{sw} < 380$ km/s 
(blue symbols), and a sample (16 events)  with $V_{sw} > 630$ km/s (red symbols). (a) the thin lines give $P_{yy}/P_{xx}$ 
for the considered events, the crosses give the average values over 16 events. (b) The crosses are the same average 
values as in 
(a); the thin solid lines give the error interval (average $\pm$ the standard deviation) for the observations 
in the fast and the slow wind. From 
about 2~Hz to 10~Hz, the difference between the fast and the slow wind is significant. 
(c) the red crosses are the average values of $P_{yy}/P_{xx}$ given in (b) for the fast wind intervals. 
(d) the blue crosses are the average values of $P_{yy}/P_{xx}$ given in (b) for the slow wind intervals. 
In c) and d), the thick black lines are the average ratios for the 2D (solid line) and isotropic (dash-dot line) distributions 
(Eqs. 2 and 3) in the same (respectively fast and slow) wind intervals. 
In (c) and (d), the thin lines on either side of the 2D and the isotropic model give the error interval of the model
(average $\pm$ the standard deviation), related to the dispersion of the spectral index. The horizontal bars 
give the frequency ranges corresponding to the proton and electron scales respectively in the fast 
and slow wind intervals.}

\label{fig5}
\end{center}
\end{figure}

To explain the variations of the observed anisotropies with the field-to-flow angle $\theta$ at a given frequency, 
different authors consider a linear combination of two of the above distributions (\citet{saur99} and references therein), 
slab, slab/R, 2D and isotropic, in different frequency domains in the solar wind. 
At frequencies [$10^{-3}$~Hz, 2 $10^{-2}$~Hz], \citet{bieber96} 
find that the turbulence in the inertial range can be 
described by $\simeq 85\%$ of the energy in 2D fluctuations, and $\simeq 15\%$ in slab fluctuations. 
\citet{leamon98} find that the proportion of slab fluctuations increases at proton scales (0.1 to 1~Hz). 
\citet{hamilton08} find an even larger proportion (more than 80\%) of slab fluctuations at proton scales. 
However, the pure 2D + slab combination is oversimplified: an angular lobe 
of wave vectors along each axis has probably to be considered. 
This is illustrated by the work of 
\citet{mangeney06} who study the variations, as a function of $\theta$, of 
the PSD of the magnetic fluctuations and of the PSD of the electric fluctuations observed by {\it Cluster} in the magnetosheath 
from 8~Hz to 2~kHz. 
They find that the electromagnetic PSD, up to  $kd_e \simeq 30$ or more, can  be explained by a 2D  ${\bf k}$-distribution, 
with an aperture angle of about  $10^{\circ}$; the electrostatic fluctuations, above $kd_e \simeq 160$ can be explained by a 
slab distribution,  with the same  aperture angle for ${\bf k}$ around ${\bf B_0}$. 

A more or less smooth transition from a 2D to an isotropic ${\bf k}$-distribution, when $f$ increases, could be a 
model of our observations more realistic than a two-components model with an increasing proportion of slab 
fluctuations when $f$ increases.

\subsection{Wavevector distribution in the fast and slow winds}

In the inertial range (resolution of about 1-minute),
\citet{dasso05} and \citet{weygand11} find  that fast 
streams are dominated by wave vectors quasi-parallel to ${\bf B_0}$, while intermediate and slow streams have 
quasi-2D ${\bf k}$-distributions. 
\citet{osman07}, using 4~s FGM data on {\it Cluster}, 
do not find any relation between the solar wind speed and the ratio of the parallel to perpendicular correlation lengths. 
\citet{hamilton08} do not find a significant difference between the  ${\bf k}$-distribution in the fast and the slow wind, 
neither in the inertial range nor at proton scales.

In our sample, the wind speed varies from 300 to 700 km/s (Fig. 1b).  
We consider separately a sample of 16 intervals with  $V_{sw} > 630$ km/s, and a sample of 16 intervals with  $V_{sw}< 380$ km/s. 
Figure 5a displays $P_{yy}/P_{xx}$ as a function of $f$ for the fast wind (red) and for the slow wind (blue). The crosses give the 
average value of each sample. We see that the average value of $P_{yy}/P_{xx}$ is larger for the fast wind, below 10~Hz. 
Is this difference significant, in view of the large dispersion of $P_{yy}/P_{xx}$ in each sample? In Figure 5b, the average 
values of Figure 5a (crosses) are shown again, as well as the error interval for each sample
(average $\pm$ the standard deviation) given by the solid lines of the same color. 
Between 2~Hz and 10~Hz, the average values given by the red (blue) crosses are at the boundary of the blue 
(red) error interval: 
the difference between  the averages of $P_{yy}/P_{xx}$ in the fast wind and the slow wind is thus significant at 
these frequencies. Figures 5c and 5d allow a comparison of the averages of $P_{yy}/P_{xx}$ observed in the fast and the 
slow wind (crosses) to $P_{yy}/P_{xx}$ 
calculated for two models of the ${\bf k}$-distribution given by \citet{saur99} (Eqs. 2 and 3).

We thus find that the 2D character of the ${\bf k}$-distribution is stronger in the fast wind than in the slow wind, 
from 1 to 10~Hz. 
This is contrary to the observations in the inertial range (quoted above). In the same way, 
simulations lead 
\citet{verdini16} to conjecture that the turbulence is more isotropic in the fast wind than 
in the slow wind, still in the inertial range. A reason for these differences could be that our frequency range is more than 
one hundred times higher than the inertial range, so that the operating physical processes can be different.

\section{Magnetic compressibility of the fluctuations}


The compressibility $P_{\parallel} / P_{\perp}$ of the 93 events of our sample (Figure 3b) increases with  the
 frequency. The average value of $P_{\parallel} / P_{\perp}$ is about 0.2 at 1~Hz, and reaches 0.5 at 50~Hz, close to the 
values displayed in Figure 1 of 
\citet{kiyani13}. Above 50~Hz, $P_{\parallel} / P_{\perp}$ is larger than the 
value 0.5 corresponding to an isotropic variance, and it still increases up to 0.65 at 200~Hz. 
To allow comparisons with other works, Figure 6a displays $P_{\parallel} / P_{3D}$, another definition of the 
compressibility, as a function of $f$. 
\citet{salem12} find that $P_{\parallel} / P_{3D}$ reaches 0.5 at 2~Hz. This is larger than what we find.

\begin{figure}[!h]
\begin{center}
\includegraphics [width=9cm]{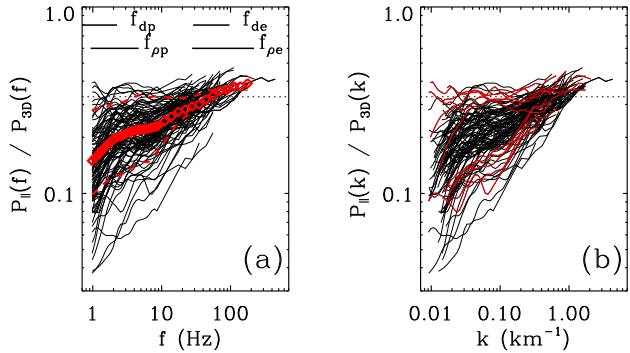}
\caption{The compressibility, here defined as $P_{\parallel} / P_{3D}$, as function of $f$ in (a) and of  
$k$ in (b) for the 93 events. 
In (a), the red diamonds give the compressibility at each frequency, averaged over the 93 events; 
the lower dashed red curve gives the average compressibility for the five events with the lowest values 
of $\beta_p$ ($0.28 < \beta_p < 0.35$), while the upper dashed red curve gives the average compressibility 
for the five events with the largest values of $\beta_p$ ($2.6 < \beta_p < 5.1$). In (b), the five lower (upper) 
red lines give the compressibility for the same five events with the lowest (largest) values of  $\beta_p$.}
\label{fig6}
\end{center}
\end{figure}

The ranges of $\beta_p$ 
values of our sample are given in Figure 1a. 
The lower red dashed line in Figure 6a is the average of $P_{\parallel} / P_{3D}$ for five events with a small $\beta_p$, and 
the upper red dashed line for five events with a large $\beta_p$ (see the selected values of $\beta_p$ in the caption). 
These red lines show that the compressibility, and the variation 
of the compressibility with $f$, depend strongly on $\beta_p$. 
 
\subsection{Compressibility in the wavenumber domain}


\begin{figure}[!h]
\begin{center}
\includegraphics [width=9cm]{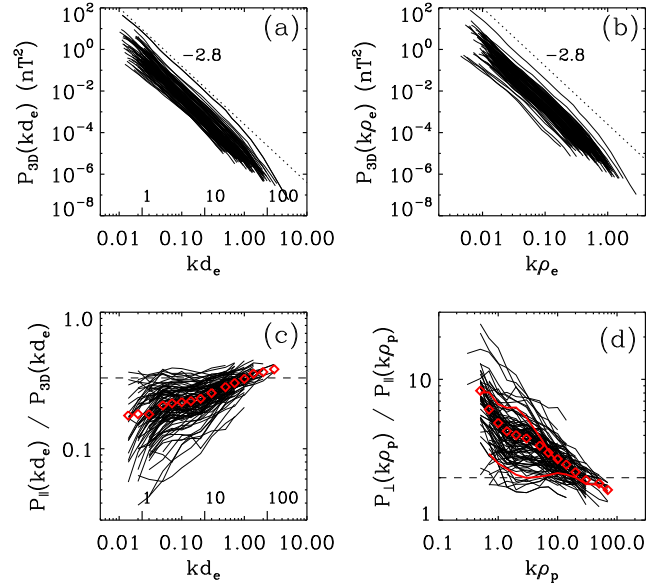}
\caption{In the wavenumber domain, for the 93 intervals. (a)  3D spectra as functions of $kd_e$.  
(b)  3D spectra as functions of $k \rho_p$. The dotted line helps to see that the spectral slopes are close to -2.8.
(c) Compressibility $P_{\parallel}/P_{3D}$ a a function of $kd_e$.
(d) Ratio  $P_{\perp}/P_{\parallel}$ as a function of $k \rho_p$. 
The upper solid red line is the average value of  $P_{\perp}/P_{\parallel}$ for the five events with the lowest values of 
$\beta_p$ ($0.28 < \beta_p < 0.35$), while the lower  solid red line is the average value for  the five events with the 
largest values of $\beta_p$ ($2.6 < \beta_p < 5.1$). In (c) and (d), the red diamonds give the average value at each scale. 
The vertical bars in the abscissae of (a) and (c) give  $kd_p=$ 1, 10 and 100. 
The horizontal dashed lines indicate values corresponding to an isotropic PSD.}
\label{fig7}
\end{center}
\end{figure}

Converting the observed frequency $f$ into a wavenumber $k = 2 \pi f /V_{sw}$ (Taylor hypothesis) we calculate the 
PSD  $P_{xx}(k)$,  $P_{yy}(k)$, $P_{zz}(k)$ and $P_{3D}(k)$ in ${\rm nT^2 km}$ (in spite of the fact that this conversion is 
problematic for anisotropic ${\bf k}$-distributions and reduced spectra; we shall return to this point in the Discussion). 
In Figure 6b, similar to Figure 6a, we see 
that the variation of the compressibility $P_{\parallel}/P_{3D}$ with the wavenumber $k$ depend strongly on $\beta_p$.  

We then normalize $k$ to the electron inertial length $d_e$, or to the gyroradii $\rho_e$ or $\rho_p$, 
and interpolate the normalized PSD at 15 values of $kd_e$, $k \rho_e$ or $k \rho_p$.
Figures 7a and 7b  give the total power $P_{3D}$ as functions of  $kd_e$ and $k \rho_e$. The slope of the spectra is close 
to  -2.8 at the largest scales. At the smallest scales, the spectra are steeper, with an exponential shape controlled by 
the electron gyroradius \citep{alexandrova12}.

Figure 7c gives  the compressibility $P_{\parallel}/P_{3D}$ as a function of $kd_e$. 
The vertical bars in the abscissae give  $kd_p=1$, 10 and 100.
With high-resolution 2D hybrid simulations of a decaying turbulence, for $\beta_p = \beta_e =$ 0.5, 
Franci  {\it et al.} (2015) obtain a compressibility 
$P_{\parallel}/P_{3D}$ increasing from 0.1, at $kd_p =$ 0.1, to 0.5 at $kd_p =$ 2, and a nearly constant (plateau) 
value $\simeq 0.6$ for $kd_p =$ 5 to 10. \citep[See also the gyrokinetic simulations in Fig. 9 of][]{tenbarge12}. 
After a kind of plateau from $kd_p = $ 2 to 10, the observed average $P_{\parallel}/P_{3D}$ (Fig. 7c) increases from 
$kd_p = $ 10 to 100, but it is always smaller than 0.4 (reached for $kd_e \simeq 2$, $kd_p =100$). We conclude that the observed 
and the simulated compressibility have similar variations up to $kd_p =10$; but the observed compressibility 
is 2 to 3 times weaker: we shall discuss this last point in section 6.
For $kd_p > 10$, a domain not reached by these simulations, the observed compressibility still increases. 


\begin{figure}[!h]
\begin{center}
\includegraphics [width=9cm]{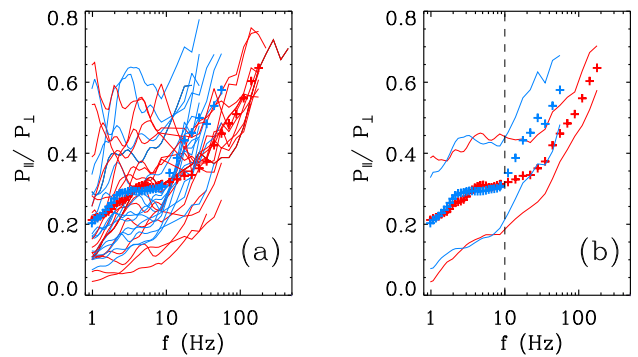}
\caption {Ratios $(P_{\parallel} / P_{\perp})(f)$ for two samples: a sample (16 events) with $V_{sw} < 380$ km/s 
(blue symbols), and a sample (16 events)  with $V_{sw} > 630$ km/s (red symbols). (a) the thin lines give 
$P_{\parallel} / P_{\perp}$ for the considered events, the crosses give the average values over 16 events. 
(b) The crosses are the same average values as in (a); the solid lines give the error interval 
(average $\pm$ the standard deviation) for the fast and the slow wind. Above about 20~Hz, 
the difference between the compressibility in the slow and the fast wind is significant. }
\label{fig8}
\end{center}
\end{figure}
\vspace{0.3cm}

\subsection{Compressibility in the fast and slow winds}

In Figure 8a, we compare the compressibility $P_{\parallel}/P_{\perp}$ for a sample of 16 intervals in the fast wind (red) 
and 16 intervals in the slow wind (blue). The crosses give the average values for each sample. In Figure 8b, taking into account 
the standard deviation of each sample, we  see that the compressibility  is the same in the fast wind and in the slow wind 
for $f < $ 10~Hz. For $f >$ 15-20~Hz, it is larger in the slow wind.
It is well-known that there is a correlation between $V_{sw}$ and $T_p$, in the solar wind: fast winds have a higher 
proton temperature than slow winds, the electron temperature having a smaller range of variation 
\citep[see {\it e.g.}][]{mangeney99}: $T_e / T_p$ is anticorrelated with $V_{sw}$ (Fig. 1b). Thus, the larger 
compressibility observed in the slow wind above 15-20~Hz could be related to larger values of $T_e / T_p$.  

\subsection{Role of $\beta_p$}

\citet{smith06} in the inertial range, \citet{hamilton08} up to 0.8~Hz, \citet{alexandrova08a} 
from 0.3~Hz to 10~Hz, find that the compressibility increases when $\beta_p$ increases. 
This is clear in Figure 9, for the present sample, at 2~Hz and 5~Hz. The slope of  $P_{\parallel}/P_{\perp}$ as a function of  
${\rm log}\beta_p$ (solid lines in Figure 9) decreases
when $f$ increases, from 5 to 22~Hz. At 56~Hz, the compressibility is no more correlated with $\beta_p$. 

\subsection{Compressibility versus different plasma parameters}

\begin{figure}[!h]
\begin{center}
\includegraphics [width=9cm]{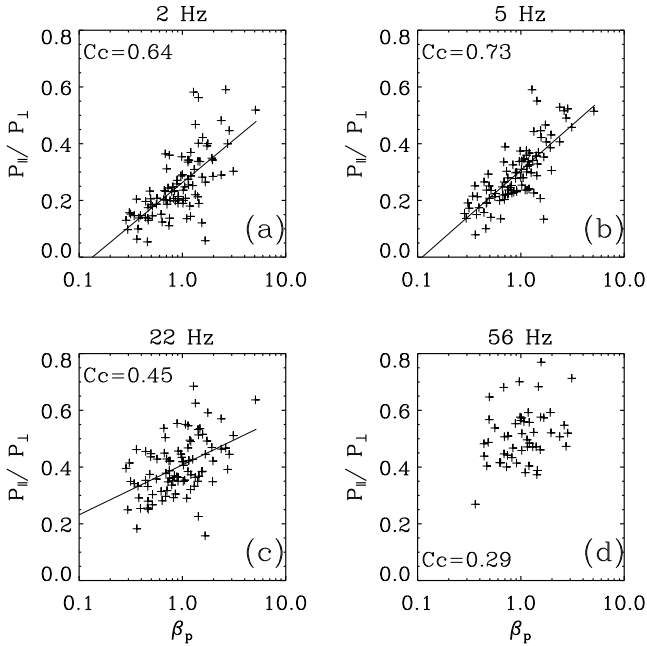}
\caption{$P_{\parallel}/P_{\perp}$ as a function of  $\beta_p$, at four frequencies (2, 5, 22 and 56~Hz), cuts of Figure 3b. 
The correlation coefficient $C_c$ between $P_{\parallel}/P_{\perp}$ and $\beta_p$ is given in each panel.}
\label{fig9}
\end{center}
\end{figure}

\begin{figure}[!h]
\begin{center}
\includegraphics [width=8.5cm]{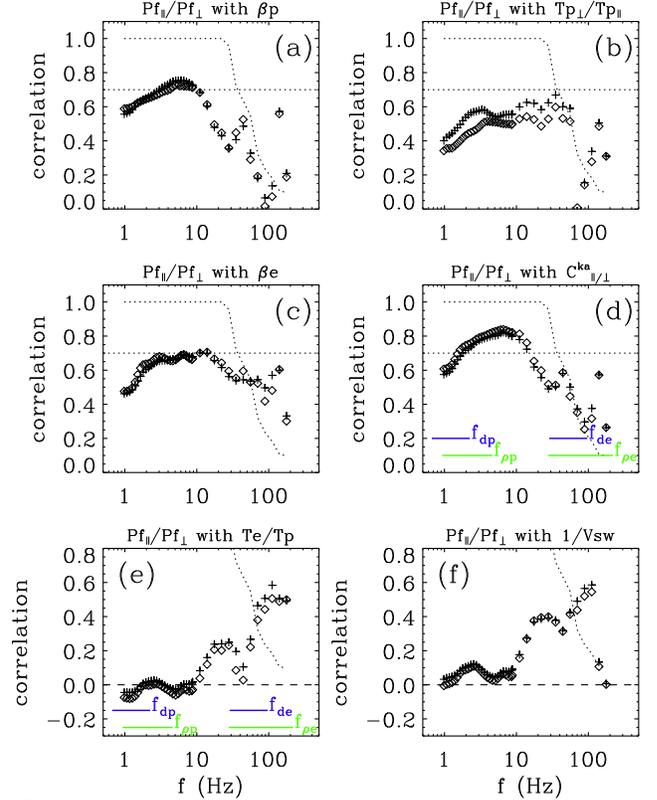}
\caption{As functions of the frequency, correlations of the observed compressibility $(P_{\parallel}/P_{\perp})(f)$ with 
plasma parameters: (a) $\beta_p$, (b) $T_{p\perp} / T_{p\parallel}$, (c) $\beta_e$, 
(d) the compressibility $C^{ka}_{\parallel/\perp}$ of kinetic Alfv\'en waves (eq. 5),
(e) $T_e/T_p$, (f) 1/$V_{sw}$ (the diamonds give the correlations  between the considered quantities, and the crosses,  
between the logarithms).
In the six panels, the dotted line gives the fraction of the 93 events which are intense enough 
at this frequency to be implied in the correlation: the fraction is 1 (93 events) up to 22~Hz, more than 
0.6 ($\simeq$ 60 events) up to 40~Hz, but less than 0.3 ($\simeq$ 30 events) above 80~Hz.  The horizontal bars in 
(d) and (e) give the frequencies corresponding to proton and electron scales as in Figs. 3 and 4.}
\label{fig10}
\end{center}
\end{figure}

The correlations of the compressibility with different plasma parameters in the solar wind 
can help to identify the nature of the dominant mode in the magnetic fluctuations. Figures 10a to 10f display 
the correlation coefficients of   $(P_{\parallel}/P_{\perp})(f)$ with local parameters related to the 
proton or electron temperature. Scatter plots of these plasma parameters are shown in Figure 1. 
The correlation of $P_{\parallel}/P_{\perp}$ with  $\beta_p$ (Fig. 10a), as a function of $f$, 
is larger than 0.7 from 3 to 10~Hz. It 
decreases clearly for $f >$ 15~Hz. The correlation with $\beta_e$ (Fig. 10c) is weaker than with $\beta_p$, but it extends 
up to 20-30~Hz.  
The correlation with  $T_{p\perp} / T_{p\parallel}$ (Fig. 10b) is good (around 0.6) from 10 to 50~Hz.
The correlation with $T_{e\perp} / T_{e\parallel}$ (not shown) is always weaker than 0.55. 

\citet{boldyrev13} \citep[see also][]{schekochihin09, tenbarge12} give analytical formulas for the 
magnetic and the electron compressibilities of sub-proton electromagnetic fluctuations, {\it i.e.} kinetic Alfv\'en waves (KAW)  
and whistler waves \citep[see also][]{gary09}. 
The whistler magnetic compressibility does not depend on $\beta$, but only on the wave vector direction. 
Conversely, the magnetic compressibility $C^{ka}_{\parallel/\perp}$ for kinetic Alfv\'en waves, 
in the framework of the electron-reduced MHD ($k \rho_p > 1$ and $k \rho_e < 1$), 
depends on $\beta_p$ and $\beta_e$, not on the wavenumber: 

\begin{equation}
C^{ka}_{\parallel/\perp} =  \frac {\beta_p + \beta_e }{2+\beta_p + \beta_e}.
\end{equation}

\noindent Figure 10d displays the correlation between the observed magnetic compressibility $P_{\parallel}/P_{\perp}$ and 
$C^{ka}_{\parallel/\perp}$, as a function of $f$. 
This correlation is everywhere larger than the correlation with $\beta_p$ (Fig. 10a). It is larger than the correlation with 
$\beta_e$ (Fig. 10c) except in a narrow frequency range around 15 to 25~Hz. 
The correlation of Figure 10d is larger than 0.7 between 1.5 and 15~Hz.

\subsection{Modes of the fluctuations}
  
\subsubsection{ For $f < $ 15-20~Hz}

At these frequencies (corresponding to $k \rho_p \leq 10$) our observations can be compared 
with the Vlasov-Maxwell results of \citet{tenbarge12}. 
These authors calculate $P_{\perp}/P_{\parallel}$ for Alfv\'en, fast and slow modes up to 
$k \rho_p = 10$, for different values of $\beta_p$, assuming that the turbulence is in critical balance 
($k_{\parallel} \propto k_{\perp}^{2/3}$) {\it i.e. } with quasi-perpendicular wave vectors. 

We have seen (Fig. 9) that $P_{\parallel}/P_{\perp}$ increases when $\beta_p$ increases: this is typical of KAW, but 
excludes the presence of fast waves for which $P_{\parallel}/P_{\perp}$ decreases when $\beta_p$ increases. 
The observations of Figure 7d are also consistent with KAW properties up to $k \rho_p = 10$ 
\citep[Fig. 2 of][]{tenbarge12}:
for the lowest values of $\beta_p$, the observed  $P_{\perp}/P_{\parallel}$ (upper solid red line) decreases strongly 
when $k \rho_p$ increases, while $P_{\perp}/P_{\parallel}$ decreases weakly  for the 
largest values of $\beta_p$ (lower solid red line). Finally, Figure 10d confirms that fluctuations with a KAW 
polarisation are generally dominant at sub-proton scales, for $f < $ 15-20~Hz). 

However, the observed $P_{\perp}/P_{\parallel}$ profiles (Fig. 7d) have a large dispersion, so that they could be 
consistent, in some cases, with other modes, oblique whistler waves with a critical balance distribution 
\citep[$k_{\parallel} \propto k_{\perp}^{1/3}$, Figure 4  of][]{tenbarge12};  or with different 
mixtures of Alfv\'en, fast and slow wave modes 
\citep[Figs. 2 and 6 of][]{tenbarge12} 
for which $P_{\perp}/P_{\parallel}$ changes strongly from $k \rho_p \simeq 0.5$ to $k \rho_p \simeq 10$. 

\subsubsection{For $f > $ 15-20~Hz}

The weaker correlation  between $P_{\perp}/P_{\parallel}$ and $C^{ka}_{\parallel/\perp}$
observed in Figure 10d above 15~Hz (and below 1.5~Hz) can be due to different 
reasons: i) Eq. 5 for $C^{ka}_{\parallel/\perp}$ 
is valid in a limited scale range, ii) the KAW are damped, and/or iii) another mode, with a different compressibility, is present. 
Let us discuss these three points: 

 i) validity of Eq. 5?  Eq. 5 is valid for $k \rho_p > 1$ and $k \rho_e < 1$. This is shown in Figure 2a of 
\citet{tenbarge12} where the compressibility is indeed equal to $C^{ka}_{\parallel/\perp}$ for 
$k_{\perp} \rho_p = $ 2 to 10, at least for the observed values $\beta_p \simeq [0.28-5.1]$ . 
In our sample, for all the events, $k \rho_p > 1$ if 
$f > $ 3~Hz, and $k \rho_e < 1$ if $f < $ 50~Hz. Thus, in Figure 10d, the weaker correlation below 3~Hz can be due to 
the lack of validity of Eq. 5. Conversely, the weaker correlation from  15-20~Hz to 50~Hz has to be due to 
other reasons.

 ii) KAW damping? Figure 10e shows the correlation between  $P_{\parallel}/P_{\perp}$ and
$T_e / T_p$ as a function of $f$. This correlation is zero up to 10~Hz, and still very small  up to 50~Hz. 
Above 80~Hz, the correlation reaches 0.5 among about 20 events (a number given by the dotted line, see the caption) 
which are the most intense at these frequencies. Figure 10f shows that the correlation between $P_{\parallel}/P_{\perp}$ and
$T_e / T_p$ (Fig. 10e) could be related to the anticorrelation of $P_{\parallel}/P_{\perp}$ with $V_{sw}$, above 20~Hz. 
According to Figure 1 of 
\citet{schreiner17} larger values of $T_e / T_p$ increase the damping of KAW, at least 
for the $\beta_p$ values of our sample, and for small scales $k_\perp \rho_p \simeq [10-100]$ 
\citep[see also][]{howes06}. The decrease of the correlation 
between $P_{\perp}/P_{\parallel}$ and $C^{ka}_{\parallel/\perp}$ above 20~Hz (Fig. 10d) can thus be due to a stronger 
damping of KAW, itself due to relatively larger values of $T_e / T_p$. Note that 
\citet{bruno15} observe that KAW (below 4~Hz) tend to disappear when the wind speed decreases. 


\begin{figure}[!h]
\begin{center}
\includegraphics [width=9cm]{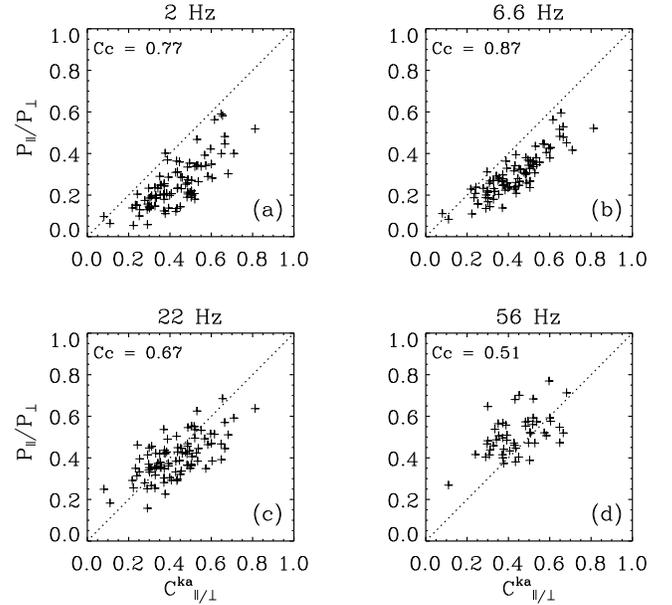}
\caption{Comparison between the compressibility $P_{\parallel}/P_{\perp}$, observed at four frequencies, and the 
compressibility of kinetic Alfv\'en waves (eq. 5). The dotted line is the bisectrix. }
\label{fig11}
\end{center}
\end{figure}

iii) another mode? The weaker correlation between $P_{\perp}/P_{\parallel}$ and $C^{ka}_{\parallel/\perp}$
for $f >$ 15-20~Hz (Fig. 10d) can be due to the presence of another mode, 
with another compressibility, superimposed or not on the KAW mode.  What could be this other mode? 
At electron scales, neither the electron temperature anisotropy (Fig. 1d) nor the heat flux are unstable: 
whistler waves are not observed.  The $\beta_e$ values are generally too low to give an electron parallel firehose 
instability \citep{Matteini_al_2013}.
The correlation of $P_{\parallel}/P_{\perp}$ with $T_e / T_p$ (Fig. 10e) above 80~Hz, which could imply the presence 
of slow  modes, is not strong. But it is consistent with the results of Figure 8, above 20~Hz: the compressibility is 
larger in the slow wind, where $T_e / T_p$ is larger.  $T_e / T_p$ varies between 0.22 and 2.2 in our sample, with 60\% of 
the intervals for $T_e / T_p < 1$ (Fig. 1b). These values seem to be too low to prevent the damping of slow-ion acoustic 
modes. However, we recall that $T_e$ and  $T_p$ are average values for the whole distributions, without distinction 
between the temperatures of core, halo, strahl or beam.  A better description of the distribution functions could 
show whether slow-ion acoustic modes could be less damped above 20~Hz, mainly in the slow wind: 
\citet{tong15} underline how the KAW damping depends on realistic distribution functions, with electron drifts.
Furthermore, the  correlation of the compressibility with $T_{p\perp} / T_{p\parallel}$ (Fig. 10b) is larger for  $f > $ 15-20~Hz 
than in the KAW frequency 
domain. According to
\citet{narita15}, quasi-perpendicular kinetic slow modes could produce a proton 
heating so that the correlation of Figure 10b could be another indication for the 
presence of kinetic slow modes, cascading above 15-20~Hz. 

In Figure 11, $P_{\parallel}/P_{\perp}$ observed at four fixed frequencies is drawn as a function of 
$C^{ka}_{\parallel/\perp}$ (Eq. 5).  
On each panel, $C_c$ gives the correlation coefficient between $P_{\parallel}/P_{\perp}$ and 
$C^{ka}_{\parallel/\perp}$. At 2~Hz the observed compressibility is always weaker than $C^{ka}_{\parallel/\perp}$, the 
compressibility expected for KAW (Fig. 11a). 
At 6.6~Hz (Fig. 11b), where $C_c$ peaks, $P_{\parallel}/P_{\perp}$ is still weaker but closer to $C^{ka}_{\parallel/\perp}$.
At 22~Hz the observed compressibility is, on an average, equal to $C^{ka}_{\parallel/\perp}$ (Fig. 11c). 
At 56~Hz, $P_{\parallel}/P_{\perp}$ is mainly larger than $C^{ka}_{\parallel/\perp}$ (Fig. 11d). 

Taking into account that the observed magnetic compressibility $(P_{\parallel}/P_{\perp})(f)$ can be underestimated 
with respect to simulations or numerical calculations (see section 6), we conclude that $P_{\parallel}/P_{\perp}$ is larger 
than the KAW compressibility for $f > $ 15-20~Hz, {\it i.e.} $k \rho_p \geq 10$. 

A few simulations of the slow-ion acoustic mode reach the scales $k \rho_p \simeq 100$ or $k \rho_e \simeq 1$ 
corresponding to our observations. 
The hybrid Vlasov-Maxwell simulations of 
\citet{valentini08} and 
\citet{valentini09} reach 
$k d_p \simeq 100$ and show the generation of ion-acoustic turbulence; but these simulations imply a temperature 
ratio $T_e/T_p = 10$  much larger than the values of our sample (Fig. 1b). 
\citet{camporeale11b} reach $k \rho_e \simeq 10$ but the electron compressibility of their simulated 
fluctuations  is weaker than the calculated electron compressibility of slow-ion acoustic waves.  An interesting result of Figure 3 
of \citet{camporeale11a}  
is that the electron compressibility of the slow-ion acoustic mode is larger for 
a small propagation angle ($\theta_{kB} = 45^{\circ}$) than for $\theta_{kB}=$ 60$^{\circ}$ or 80$^{\circ}$. This is 
another point in favor of the presence of slow-ion acoustic modes in our data. Indeed, we have concluded (section 4.2) that the 
${\bf k}$-distribution is more isotropic, 
so that small $\theta_{kB}$ angles  are more frequent, when $f$ increases. Thus, a larger electron compressibility should 
be found when $f$ increases ($\theta_{kB}$ decreases) if slow-ion acoustic modes are present. 
We did not measure the {\it electron} compressibility, but  Figures 6a and 11 show that the {\it magnetic}  
compressibility increases when $f$ increases. Calculations similar to those of 
\citet{gary09} would tell us whether the  {\it magnetic}  compressibility of slow-ion acoustic modes 
increases when $\theta_{kB}$ decreases. 

\section{Discussion}

In this work, we consider the reduced power spectral densities  
$P_{xx}(f)$,  $P_{yy}(f)$ and $P_{zz}(f)$ of the magnetic field fluctuations observed  
at a given frequency $f$ in the three directions with respect to the quasi-local magnetic field ${\bf B_0}$ (4~s average),  
$P_{\parallel}(f) = P_{zz}(f)$ and $P_{\perp}(f) =  P_{xx}(f) + P_{yy}(f)$. 
The sampling direction is in the plane ({\bf x}, {\bf z}) {\it i.e.} perpendicular to the {\bf y} direction. 

In section 5, the magnetic compressibility observed in the frequency domain, $(P_{\parallel}/P_{\perp})(f)$ is compared with 
analytical and numerical calculations and simulations in the wave vector domain, which only consider two components 
$k_{\parallel}$ and $k_{\perp}$, and give the compressibility $(P_{\parallel}/P_{\perp})(k_{\perp})$. 
But this comparison is not really valid: we cannot assume that $(P_{\parallel}/P_{\perp})(f)$ is equal to 
$(P_{\parallel}/P_{\perp})(k_{\perp})$. Indeed, if there is only one wave vector ${\bf k_1}$ perpendicular to ${\bf B_0}$, 
along ${\bf x}$ for instance, the PSD in the directions perpendicular to ${\bf k_1}$, 
$P_{yy}(k_1)$ and $P_{zz}(k_1)$, will be observed as  $P_{yy}(f_1)$ and $P_{zz}(f_1)$ at the same frequency 
$f_1 = {\bf k_1} \cdot {\bf V_{sw}} / 2 \pi $ 
\citep[see][section 2.1]{podesta12},  and will display a phase relation 
between $\delta B_y$ and $\delta B_z$ corresponding to their polarisation. 
If there is also a wave vector ${\bf k_2}$ perpendicular to ${\bf B_0}$, along ${\bf y}$, with a wavenumber $k_2 = k_1$, 
$P_{xx}(k_2)$ and $P_{zz}(k_2)$ will be observed as  $P_{xx}(f_2)$ and $P_{zz}(f_2)$ at the same frequency 
$f_2 = {\bf k_2} \cdot {\bf V_{sw}} / 2 \pi $, 
but  $f_2$ will be much smaller than $f_1$ because ${\bf k_2}$ is more or less perpendicular to the sampling direction 
${\bf V_{sw}}$. Thus, even if two wave vectors 
perpendicular to ${\bf B_0}$ have the same $k_{\perp}$, they will not be observed at the same frequency in  
$P_{xx}(f)$,  $P_{yy}(f)$ and $P_{zz}(f)$. 
Inversely, at a given frequency,  $P_{xx}(f)$,  $P_{yy}(f)$ and $P_{zz}(f)$ do not correspond to the same wavenumbers 
and wave vectors: 
they do not belong to the same fluctuations, there is no clear phase relation between the components $x$, $y$ and $z$.  
 Furthermore, a PSD like $P_{zz}$,  or like $P_{\perp} = P_{xx} + P_{yy}$ or $P_{3D} = P_{\perp} + P_{zz}$,
at a given frequency, is a mixture of fluctuations with different wavenumbers and different phases. 
Thus, analytical and 2D numerical calculations, which involve the same wave vectors for all the PSD, and simulations 
which do not "fly through" the simulation box with the solar wind velocity, 
cannot predict exactly what is observed at a given frequency. This is probably one of the reasons why the observed 
compressibility is sometimes weaker than the simulated or calculated compressibility (see section 5). 

The Taylor hypothesis has been assumed to be valid from 1 to 100~Hz: in Figures 6b and 7, the wavenumber 
dependency relies on this hypothesis. 
However, at proton or electron scales, some fluctuations can be present, with a frequency $\omega_0$ in the solar wind 
frame non-negligible with respect to  ${\bf k} . {\bf V_{sw}}$. 
Thus, any detailed comparison of observations with calculations in the wavenumber domain (section 5.1) 
has to take into account 
the possibility that $k$ is not always proportional to $\omega$, in the kinetic range, and that wave modes could be present 
with  a non-negligible $\omega_0$. What could be these wave modes?

\citet{howes14} show that the Taylor hypothesis is valid for all the linear modes (Alfv\'en, fast, slow) in the 
near-Earth solar wind, except for quasi-parallel whistler waves. Our frequency range [1 - 200~Hz] is below the 
electron gyrofrequency range $f_{ce}$: in our sample, $f_{ce}$ varies from 150 to 550~Hz, 
with an average value of 300~Hz. Quasi-parallel whistler waves could thus be present below $f_{ce}$ 
\citep{lacombe14, stansby16,Kajdic16}; 
but their characteristic right-handed polarisation with respect to ${\bf B_0}$,  
and the corresponding spectral bump, are not observed, in our sample; so that even if weak 
underlying quasi-parallel whistlers were present, they could not play an important role. 

The observed frequency range is well above the proton gyrofrequency range $f_{cp} = $  [0.08 - 0.3~Hz]. 
Thus, ion Bernstein modes, at low harmonics (1, 2 or 3) of $f_{cp}$ cannot be observed. 

Our frequency range is well below the proton plasma frequency ($f_{pi}$ varies from 400~Hz to 1.5~kHz, 
average $\simeq$~800~Hz): slow-ion acoustic waves could be present, which propagate below $f_{pi}$. 
But the only magnetic compressibility cannot allow the identification of 
non-Alfv\'enic wave modes \citep{narita15}. 
This is illustrated by \citet{tenbarge12}. 
To remove these ambiguities, fluctuations other than the magnetic field fluctuations, or other  transport ratios, would 
be considered, notwithstanding the fact that each wave identifier has its own ambiguity 
\citep{krauss94, denton98, klein12}. 
However, at small scales ($f >$ 15-20~Hz, see section 5.5.2),  the observed compressibility is larger than the KAW 
compressibility (Fig. 11). It is larger in the slow wind than in the fast wind (Fig. 8) {\it i.e.} larger when 
the electron to proton temperature ratio $T_e / T_p$ is larger (Fig. 1b).  
Is there a mode which has a compressibility larger 
than the KAW compressibility, which could be more present in the slow wind, where  $T_e / T_p$ is larger, and which can 
propagate at oblique or small angles with respect to $B_0$? According to Figure 1 of \citet{gary92}, 
if $T_e/T_p \geq 1$ and $\beta_p > 1$, there is always a slow mode which is only weakly 
damped for small propagation angles. This is valid for MHD modes, and up to $k d_p$ = 0.1 
\citep{gary-win92, krauss94}. This is still valid up to $k \rho_e \simeq 10$ 
\citep{camporeale11a} at least for $\beta_p = \beta_e = 0.5$. The magnetic compressibility of this kinetic mode 
has to be calculated to be compared to our observations. 

The comparison of the observed polarisation ratios with those of linear modes is not always justified. 
But it could be more valid at the smallest scales ($f >$ 20~Hz, or $k \rho_e > $ 0.2),
when some angles $\theta_{kB}$ are probably smaller than the values giving the critical balance condition, 
{\it i.e.} smaller than the values at which the role of non-linear effects becomes dominant \citep[Fig. 1 of][]{howes08}. 

\section{Summary and conclusion}

We have analyzed 93 10-minute intervals of the solar wind magnetic field fluctuations observed on {\it Cluster}~1, 
from 1~Hz to about 
200~Hz. These intervals, between 2001 and 2005, are selected with two conditions: 1) there is no indication of a 
connection to the Earth's bow shock; as a consequence, the {\it Cluster} orbit implies that the field-to-flow angle 
$\theta$ is large; 2) there is no indication of the presence of quasi-parallel whistler waves, which are 
easily detected by their quasi-circular right-handed polarisation \citep{lacombe14}. 

The power spectral density (PSD) of the magnetic field turbulence is measured in the three directions of the magnetic field 
aligned (MFA) frame, an average quasi-local frame given by the 4~s magnetic field data. The PSD is anisotropic. It 
decreases when $f$ increases; it goes down to the instrumental background noise at different frequencies 
in the different directions.  As explained in section~3 and Appendix, we only consider  the PSD when it is larger 
than 3 times the noise in any direction; then, we subtract the noise to obtain the corrected signal.

The corrected power spectral density $P(f)$ is itself anisotropic. It is maximum in the ${\bf y}$ direction, which 
is perpendicular to ${\bf B_0}$ and to the radial direction.
$P(f)$ is weaker in the ${\bf x}$ direction, perpendicular to ${\bf B_0}$ and ${\bf y}$: 
the ratio $P_{yy}(f)/ P_{xx}(f) > 1$ implies that the transverse fluctuations are not gyrotropic at a given frequency $f$ in the 
spacecraft frame. 

Following \citet{saur99}, 
the non gyrotropic ratio $(P_{yy}/ P_{xx})(f)$ gives indications on the shape of a gyrotropic distribution of ${\bf k}$. 
We find that the ${\bf k}$-distribution $P({\bf k})$ is mainly 2D 
({\it i.e.} $k_{\perp} \gg k_{\parallel}$) 
up to about 6~Hz, {\it i.e.} $k d_e \simeq 0.1$ or $k \rho_p \simeq 3$. 
There is also a slight proportion of wave vectors parallel to ${\bf B_0}$. 
By a separate study of intervals of fast wind and of slow wind, we show that $P({\bf k})$ is closer to a 2D distribution 
in the fast wind, and contains a larger proportion of wave vectors parallel to ${\bf B_0}$ in the slow wind, at $f < $ 10~Hz. 
At frequencies $f>$ 10~Hz, $P({\bf k})$ is less 2D, and tends towards isotropy around 50~Hz: wave vectors oblique or 
parallel to ${\bf B_0}$ become relatively more important when the frequency increases (section 4.2).  

We call compressibility the anisotropy ratio $(P_{\parallel} / P_{\perp})(f)$ between the compressive and the transverse 
fluctuations: this polarisation ratio increases from 1 to 200~Hz.  
From 2~Hz to about 15-20~Hz, the observed compressibility is very well correlated (Fig. 10d) and nearly equal (Fig. 11) to 
the compressibility (Eq. 5) of linear kinetic Alfv\'en waves (KAW), which only depends on the beta factor 
$\beta = \beta_p + \beta_e$: 
fluctuations with a KAW-like polarisation are 
dominant below 20~Hz, {\it i.e.} below $k d_e \simeq 0.2$ or $k \rho_p \simeq 10$.

At smaller scales, $f>$ 20~Hz, we suggest that the fluctuations could be related to a slow-ion acoustic mode. 
The identification of the three modes, Alfv\'enic or compressive, and the estimation of their damping rates and of their 
polarisation and transport ratios is a difficult task in the kinetic range  where the modes cross each other in the 
dispersion plane \citep{krauss94, klein15}. 
Calculations of the dispersion relation and of the compressibility 
of the modes, for different propagation directions, different $\beta_p$ and different $T_e/T_p$, would be necessary 
to confirm (or infirm) the presence of slow-ion acoustic modes 
with a large magnetic compressibility and relatively small propagation angles, for $k \rho_p \geq 10$. 

Note, however, that recent studies \citep{Chen_Boldyrev_2017, Passot_al_2017} suggest that the 
compressibility of KAW can further increase when reaching the electron inertial length scale, 
thus in possible good agreement with the trend observed in this dataset for frequencies above 20 Hz.

\begin{acknowledgments}
We thank Nicole Cornilleau-Wehrlin and Patrick Canu, PIs of the STAFF experiment, for the use of the STAFF-SC and STAFF-SA data.
We thank Pierrette DŽcrŽau for the WHISPER data. The FGM data (PIs: A. Balogh and E. Lucek) and the CIS data (PIs: H. Rme and I. Dandouras) come from the Cluster Science Data System (ESA). The PEACE data (PI: A. Fazakerley) are taken from the Cluster Science Archiv (ESA). The French contribution to the Cluster project has been supported by the European Space Agency (ESA) and by the Centre National d'Etudes Spatiales (CNES).
LM was supported by the UK STFC grant ST/N000692/1.
\end{acknowledgments}

\appendix
\section{Subtraction of the background noise}\label{appendix}

For a precise measurement of weak spectral levels on {\it Cluster}, the subtraction of the instrument noise 
has been justified by \citet{alexandrova10}: 
the frequency-dependent background instrument noise is a random variable; the solar wind turbulence is a random 
variable; these random variables are independent. It is well known that the average value of the sum of two independent 
random variables is the sum of their average values. Thus, 
the true spectral level at the frequency $f$ is the difference $P_{3D}(f)-bn_{3D}(f)$, where 
$P_{3D}(f)$ is the average of the trace of the spectral matrix measured on the triaxial search coils (signal + noise), and 
$bn_{3D}(f)$ is the average of the trace of the background noise (additive noise). 
The background noise is generally called sensitivity, but this name is misleading because it implies a threshold, 
which has to be crossed, not an additive noise which has to be subtracted. 


\begin{figure}[!h]
\begin{center}
\includegraphics [width=9cm]{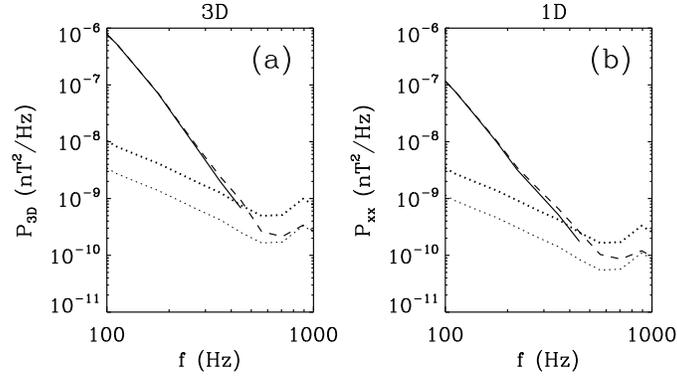}
\caption{Example of subtraction of the background noise on {\it Cluster}~1 (day 2004/01/22, 01:55-02:05 UT) 
(a) for the 3D spectrum $P_{3D}(f)$, (b) for $P_{xx}(f)$. On each panel, the dashed line gives the measured spectrum 
(signal + noise), the solid line gives the corrected spectrum, at frequencies where the conditions (6) and (7) are both fulfilled. 
The lowest dotted line gives the background noise, respectively $bn_{3D}(f)$ ( left panel) 
and $bn_{1D}(f)$ (right panel). The upper dotted line on each panel gives 3 times the noise, the lowest 
significant value of the measured signal (Eqs. \ref{App1} and \ref{App2}).}
\label{fig 12}
\end{center}
\end{figure}

The average background noise is easily measured every year in Summer, when {\it Cluster} 
crosses the magnetospheric lobes: 
in these regions, the {\it in situ} fluctuations are so weak that the measured signal is considered as the receiver noise. 
This instrument noise is close to the values obtained by Earth's ground measurements \citep{cornilleau03}. 
\citet{robert14} show, in their 
Figure 9, that the background noise is rather stable from 2001 to 2004, and even until 2011. 
At the STAFF-SC frequencies (below 8~Hz) the observed solar wind spectra (Figure 2) are intense enough, so that there 
is no need to subtract the noise. On STAFF-SA (above 8~Hz), from 2001 to 
2005, we consider that the 3D-background noise $bn_{3D}(f)$ is given by a one-hour average of 
the 3D power $P_{lobe}$ measured in a lobe on 2004/06/03 (14:00-15:00 UT) on  {\it Cluster}~1. 
Histograms of ${\rm log} P_{lobe}(f)$ during this interval, at different $f$, show that the instantaneous background noise 
can be as large as 2 to 3 times the average value $bn_{3D}(f)$ below 60~Hz, and 1.3 to 1.5 times $bn_{3D}(f)$ above 60~Hz.  
(Above 60~Hz, the power of each STAFF-SA spectrum relies on 8 times more measures than below 60~Hz, so that the relative 
uncertainty on the power level is weaker). 
To reach the true spectral level by a subtraction $P_{3D}(f)-bn_{3D}$, we have to take into account 
this uncertainty on $bn_{3D}$, which is at most 3 times $bn_{3D}$. The corrected spectral level 
is the average value of $P_{3D}-bn_{3D}$, but at the only frequencies where $P_{3D}$ is larger than $3 bn_{3D}$,

\begin{equation}\label{App1}
P_{3D}(f) \geq 3 bn_{3D}(f).
\end{equation} 
\noindent There is no simple reliable way to 
separate the signal and the instrument noise at frequencies for which $P_{3D}$ is weaker than 3$bn_{3D}$.

For the background noise in every direction $x$, $y$ and $z$, we use $bn_{1D}(f) = bn_{3D}(f)/3 $ (as in Figure 2). 
Even if the 
three search coils in the spacecraft frame (SR2) have not exactly the same 1D-background noise, it is not 
useful to distinguish them: indeed, the on-board (SR2) spectral matrix (signal + noise) is projected and analyzed in the 
Magnetic Field Aligned frame, $x$, $y$ and $z$, so that the role played by each search coil is time-dependent, through 
the angles between the SR2 and the MFA frames. Anyway, the considered uncertainty (3 times the noise) broadly 
takes into account the differences between the three search coils. 

The present work is based upon measurements of the PSD anisotropies.
We have to subtract the 1D background noise from $P_{xx}(f)$, $P_{yy}(f)$ and $P_{zz}(f)$. As 
$P_{xx}(f)$ (dashed lines in Fig. 2) is the weakest 1D spectrum (above 8~Hz), we first have to check that the noise 
subtraction is significant for$P_{xx}$: indeed, the subtraction from $P_{3D}(f)$ can  
be valid (if $P_{3D}(f)$ is larger than $3 bn_{3D}(f)$), while the subtraction from $P_{xx}(f)$ is not valid, if 
$P_{xx}(f)$ is equal to the 1D background noise. 
To prevent errors in the subtraction of the noise from $P_{xx}(f)$ we thus put a second condition: 
\begin{equation}\label{App2}
P_{xx}(f) \geq 3 bn_{1D}(f).
\end{equation}

\noindent  As $P_{xx}(f)$ is always weaker than $P_{yy}(f)$ and $P_{zz}(f)$, this second condition implies that the 1D noise 
can be subtracted without errors in every direction $x$, $y$ and $z$.
Figure 12 illustrates the two conditions (Eqs. \ref{App1} and \ref{App2}) to get valid measurements of a 3D spectrum, of $P_{xx}(f)$, 
$P_{yy}(f)$ and $P_{zz}(f)$, 
and thus of the anisotropy ratios of Figure 3. The lower dotted lines in Figure 12 give the background noise for the 3D noise 
(left panel) and the 1D noise (right panel). The upper dotted lines respectively give $3 bn_{3D}$ and $3 bn_{1D}$. 
The dashed lines give $P_{3D}$ (left panel) and $P_{xx}$ (right panel) before any subtraction. The solid lines give the 
corrected spectra, which obey the two above conditions, and after subtraction respectively of $bn_{3D}$ and $bn_{1D}$.

All the spectra used in the present paper are corrected in this way.
\\
\\

\bibliographystyle{apj.bst}

\end{document}